\documentclass{hep99}

\def\lqcd{\Lambda_{\rm QCD}}
\begin{document}

\input epsf.sty

\title{$B$ Decays in the Upsilon Expansion}

\author{Zoltan Ligeti}

\address{Theory Group, Fermilab, P.O.\ Box 500, Batavia, IL 60510, USA \\[3pt]
E-mail: {\tt ligeti@fnal.gov} %% \hfil FERMILAB-Conf-99/336-T
}

\abstract{Theoretical predictions for $B$ decay rates are rewritten in terms of
the Upsilon meson mass instead of the $b$ quark mass, using a modified
perturbation expansion.  The theoretical consistency is shown both at low and
high orders.  This method improves the behavior of the perturbation series for
inclusive and exclusive decay rates, and the largest theoretical error in the
predictions coming from the uncertainty in the quark mass is eliminated. 
Applications to the determination of CKM matrix elements, moments of inclusive
decay distributions, and the $\bar B\to X_s\gamma$ photon spectrum are
discussed.} 

\maketitle

\section{Introduction}

Testing the Cabibbo--Kobayashi--Maskawa (CKM) description of quark mixing and
$CP$ violation is a large part of the high energy experimental program in the
near future.  The goal is to overconstrain the unitarity triangle by directly
measuring its sides and (some) angles in several decay modes.  If the value of
$\sin2\beta$, the $CP$ asymmetry in $B\to \psi K_S$, is near the CDF central
value~\cite{sin2beta}, then searching for new physics will require precise
measurements of the magnitudes and phases of the CKM matrix elements. 
Inclusive $B$ decay rates can give information on $|V_{cb}|$, $|V_{ub}|$,
$|V_{ts}|$, and $|V_{td}|$.

The theoretical reliability of inclusive measurements can be competitive with
the exclusive ones.  For example, for the determination of $|V_{cb}|$ model
dependence enters at the same order of $\Lambda_{\rm QCD}^2/m_{c,b}^2$
corrections from both the inclusive semileptonic $\bar B\to X_c e\bar\nu$ width
and the $\bar B\to D^* e\bar\nu$ rate near zero recoil.  It is then important
to test the theoretical ingredients of these analyses via other measurements.

The main uncertainty in theoretical predictions for inclusive $B$ decay rates
arise from the poorly known quark masses which define the phase space, and the
bad behavior of the series of perturbative corrections when it is written in
terms of the pole mass.  Only the product of these quantities, the decay
widths, are well-defined physical quantities; while perturbative multi-loop
calculations are most conveniently done in terms of the pole mass.  Of course,
one would like to eliminate any quark mass from the predictions in favor of
physical observables.  Here we present a new method of eliminating $m_b$ in
terms of the $\Upsilon(1S)$ meson mass~\cite{upsexp}.

\section{Upsilon Expansion}

Let us consider, for example, the inclusive $\bar B\to X_u e\bar\nu$ decay 
rate.  At the scale $\mu = m_b$,
\begin{eqnarray}\label{bupole}
 \Gamma(B\to X_u e\bar\nu) = {G_F^2 |V_{ub}|^2\over 192\pi^3}\, m_b^5\,
  \bigg[ 1 - 2.41 {\alpha_s\over\pi}\, \epsilon  \nonumber\\
 - 3.22 {\alpha_s^2\over\pi^2} \beta_0\, \epsilon^2 - \ldots
   - {9\lambda_2 - \lambda_1 \over 2m_b^2} + \ldots \bigg] . 
\end{eqnarray}
The variable $\epsilon \equiv 1$ denotes the order in the modified expansion,
and $\beta_0=11-2n_f/3$.  In comparison, the expansion of the $\Upsilon(1S)$ 
mass in terms of $m_b$ has a different structure,
\begin{equation}\label{upsmass}
m_\Upsilon = 2m_b \bigg\{\! 1 - {(\alpha_s C_F)^2\over8} \bigg[ \epsilon 
  + {\alpha_s\over\pi} \bigg(\! \ell + \frac{11}6 \!\bigg) \beta_0 \epsilon^2 
  + \ldots \bigg]\! \bigg\} , 
\end{equation}
where $\ell=\ln[\mu/(m_b\alpha_s C_F)]$ and $C_F=4/3$.  In this expansion we
assigned to each term one less power of $\epsilon$ than the power of
$\alpha_s$.  It is also convenient to choose the same renormalization scale
$\mu$.  The prescription of counting $\alpha_s^n$ in $B$ decay rates as order
$\epsilon^n$, and $\alpha_s^n$ in $m_\Upsilon$ as order $\epsilon^{n-1}$ is the
upsilon expansion.  It combines different orders in the $\alpha_s$ perturbation
series in Eqs.~(\ref{bupole}) and (\ref{upsmass}), but as it is sketched below,
this is the consistent way of combining these expressions.

At large orders in perturbation theory, the coefficient of $\alpha_s^n$ in
Eq.~(\ref{bupole}) has a part which grows as $C n! \beta_0^{n-1}$.  For large
$n$, this divergence is cancelled by the $\alpha_s^{n+1}$ term in 
Eq.~(\ref{upsmass}), whose coefficient behaves as $(1/\alpha_s) (C/5) n!
\beta_0^{n-1}$~[3--5].  %% \cite{ugo,andre,Beneke}.  
The crucial $1/\alpha_s$ factor arises
because the coefficient of $\alpha_s^{n+1}$ in Eq.~(\ref{upsmass}) contains a
series of the form $(\ell^{n-1}+ \ell^{n-2}+ \ldots+1)$ which exponentiates for
large $n$~to~give~$\exp(\ell) = \mu/(m_b\alpha_s C_F)$, and corrects the
mismatch of the power of $\alpha_s$ between the two series. 

The infrared sensitivity of Feynman diagrams can be studied by introducing a
fictitious infrared cutoff $\lambda$.  The infrared sensitive terms are
non-analytic in $\lambda^2$, such as $(\lambda^2)^{n/2}$ or
$\lambda^{2n}\ln\lambda^2$, and arise from the low momentum part of Feynman
diagrams.  Linear infrared sensitivity (terms of order $\sqrt{\lambda^2}$) are
a signal of $\Lambda_{\rm QCD}$ effects, quadratic sensitivity (terms of order
$\lambda^2\ln \lambda^2$) are a signal of $\Lambda_{\rm QCD}^2$ effects, etc. 
From Refs.~\cite{Beneke,SAZ} it follows that the linear infrared sensitivity
cancels in the upsilon expansion to order $\epsilon^2$ (probably to all orders
as well, but the demonstration of this appears highly non-trivial).
 
Thus, the upsilon expansion is theoretically consistent both at large orders
for the terms containing the highest possible power of $\beta_0$, and to order
$\epsilon^2$ including non-Abelian contributions.  

The most important uncertainty in this approach is the size of nonperturbative
contributions to $m_\Upsilon$ other than those which can be absorbed into
$m_b$.  If the mass of heavy quarkonia can be computed in an operator product
expansion then this correction is of order $\lqcd^4 /(\alpha_s m_b)^3$ by
dimensional analysis.  Quantitative estimates, however, vary in a large range,
and it is preferable to constrain such effects from data.  We use $100\,$MeV to
indicate the corresponding uncertainty.  Finally, if the nonperturbative
contribution to $\Upsilon$ mass, $\Delta_\Upsilon$, were known, it could be
included by replacing $m_\Upsilon$ by $m_\Upsilon-\Delta_\Upsilon$ on the left
hand side of Eq.~(\ref{upsmass}).  

There are three surprising facts that are either accidental or indicate that
the nonperturbative contributions may be small: 1) applications in terms of the
$\Upsilon(2S)$ mass give consistent results~\cite{upsexp}; 2) the $D\to Xe\nu$
rate in terms of the $\psi$ mass works (un)reasonably well~\cite{upsexp}; 3)
sum rule calculations for $e^+e^-\to b\bar b$ find that the 1S $b$ quark mass
(defined as half of the perturbative part of $m_{\Upsilon(1S)}$) is only
$20\,$MeV different from $m_{\Upsilon(1S)}/2$~\cite{1Smass}.

\section{Application}

Substituting Eq.~(\ref{upsmass}) into Eq.~(\ref{bupole}) and collecting terms
of a given order in $\epsilon$ gives
\begin{eqnarray}\label{buups}
\Gamma(\bar B\to X_u e\bar\nu) &=& {G_F^2 |V_{ub}|^2\over 192\pi^3}\,
  \bigg({m_\Upsilon\over2}\bigg)^5 \\ 
&&\times \Big[ 1 - 0.115\epsilon - 0.031 \epsilon^2 
  - \ldots \Big] . \nonumber
\end{eqnarray}
The complete order $\alpha_s^2$ result calculated recently~\cite{TvR} is
included.  Keeping only the part proportional to $\beta_0$, the coefficient of
$\epsilon^2$ would be $-0.035$.  The perturbation series, $1 - 0.115\epsilon -
0.031\epsilon^2$, is better behaved than the series in terms of the $b$
quark pole mass, $1 - 0.17\epsilon - 0.10\epsilon^2$, or the series expressed
in terms of the $\overline{\rm MS}$ mass, $1 + 0.30\epsilon + 0.13\epsilon^2$. 
The uncertainty in the decay rate using Eq.~(\ref{buups}) is much smaller than
that in Eq.~(\ref{bupole}), both because the perturbation series is better
behaved, and because $m_\Upsilon$ is better known (and better defined) than
$m_b$.  The relation between $|V_{ub}|$ and the total semileptonic $\bar B\to
X_u e\bar\nu$ decay rate is~\cite{upsexp}
\begin{eqnarray}\label{Vub}
|V_{ub}| &=& (3.04 \pm 0.06 \pm 0.08) \times 10^{-3} \nonumber\\
&&\times \left( {{\cal B}(\bar B\to X_u e\bar\nu)\over 0.001}
  {1.6\,{\rm ps}\over\tau_B} \right)^{1/2} ,
\end{eqnarray}
The first error is obtained by assigning an uncertainty in Eq.~(\ref{buups})
equal to the value of the $\epsilon^2$ term and the second is from assuming a
$100\,$MeV uncertainty in Eq.~(\ref{upsmass}).  The scale dependence of
$|V_{ub}|$ due to varying $\mu$ in the range $m_b/2< \mu <2m_b$ is less than
1\%.  The uncertainty in $\lambda_1$ makes a negligible contribution to the
total error.  Although ${\cal B}(\bar B\to X_u e\bar\nu)$ cannot be measured
without significant experimental cuts, for example, on the hadronic invariant
mass, this method will reduce the uncertainties in such analyses as well.

The $\bar B\to X_c e\bar\nu$ decay depends on both $m_b$ and $m_c$.  It is 
convenient to express the decay rate in terms of $m_\Upsilon$ and $\lambda_1$
instead of $m_b$ and $m_c$, using Eq.~(\ref{upsmass}) and
\begin{equation}\label{mbmc}
m_b - m_c = \overline{m}_B - \overline{m}_D + \bigg( 
  {\lambda_1\over 2\overline{m}_B} - {\lambda_1\over 2\overline{m}_D} \bigg) 
  + \ldots \,,
\end{equation}
where $\overline{m}_B = (3m_{B^*}+m_B)/4=5.313\,$GeV and $\overline{m}_D =
(3m_{D^*}+m_D)/4=1.973\,$GeV.  We then find

\begin{eqnarray}\label{bcups}
\Gamma(\bar B\to X_c e\bar\nu) &=& {G_F^2 |V_{cb}|^2\over 192\pi^3}
  \bigg({m_\Upsilon\over2}\bigg)^5\, 0.533 \\
&&\times \Big[ 1 - 0.096\epsilon - 0.029_{\rm BLM}\epsilon^2 \Big] , \nonumber
\end{eqnarray}
where the phase space factor has also been expanded in $\epsilon$, and the
BLM~\cite{BLM} subscript indicates that only the corrections proportional to
$\beta_0$ have been kept.  For comparison, the perturbation series in this
relation written in terms of the pole mass is $1- 0.12\epsilon-
0.07_{\rm BLM}\epsilon^2$~\cite{LSW}.  Including the terms proportional to
$\lambda_{1,2}$, Eq.~(\ref{bcups}) implies~\cite{upsexp}
\begin{eqnarray}\label{Vcb}
|V_{cb}| &=& (41.6 \pm 0.8 \pm 0.7 \pm 0.5) \times 10^{-3} \nonumber\\
&&\times \eta_{\rm QED} \left( {{\cal B}(\bar B\to X_c e\bar\nu)\over0.105}\,
  {1.6\,{\rm ps}\over\tau_B}\right)^{1/2} ,
\end{eqnarray}
where $\eta_{\rm QED}\sim1.007$ is the electromagnetic radiative correction. 
The uncertainties come from assuming an error in Eq.~(\ref{bcups}) equal to the
$\epsilon^2$ term, a $0.25\,{\rm GeV}^2$ error in $\lambda_1$, and a $100\,$MeV
error in Eq.~(\ref{upsmass}), respectively.  The second uncertainty can be
reduced by determining $\lambda_1$ (see below).

The theoretical uncertainty hardest to quantify in the predictions for
inclusive $B$ decays is the size of quark-hadron duality violation.  This was
neglected in Eqs.~(\ref{Vub}) and (\ref{Vcb}).  It is believed to be
exponentially suppressed in the $m_b\to\infty$ limit, but its size is poorly
known for the physical $b$ quark mass.  Studying the shapes of inclusive decay
distributions is the best way to constrain this experimentally, since duality
violation would probably show up in a comparison of different spectra.  The
shape of the lepton
energy~[11--14] %% \cite{gremmetal,gremmetal2,Volo,GK} 
and hadron invariant
mass~\cite{FLSmass1,FLSmass2,GK} spectra in semileptonic $\bar B\to X_c e
\bar\nu$ decay, and the photon spectrum in $\bar B\to
X_s\gamma$~[17--21] %% \cite{AZ,LLMW,Bauer,KaNe,AI} 
can also be used to determine the heavy quark
effective theory (HQET) parameters $\bar\Lambda$ and $\lambda_1$.  The extent
to which these determinations agree with one another will indicate at
what level to trust predictions for inclusive rates.  

Last year CLEO measured the first two moments of the hadronic invariant
mass-squared ($s_H$) distribution, $\langle s_H - \overline{m}_D^2 \rangle$ and
$\langle (s_H - \overline{m}_D^2)^2 \rangle$, subject to the constraint $E_e >
1.5\,$GeV~\cite{CLEOparams}.  Here $\overline{m}_D = (m_D+3m_{D^*})/4$.  Each
of these measurements give an allowed band on the $\bar\Lambda - \lambda_1$
plane.  Their intersection gives (at order $\alpha_s$)~\cite{CLEOparams}
\begin{eqnarray}\label{cleonums}
  \bar\Lambda &=& (0.33 \pm 0.08)\, {\rm GeV} , \nonumber\\
  \lambda_1 &=& -(0.13 \pm 0.06)\,{\rm GeV}^2 .  
\end{eqnarray}
This agrees well with the analysis of the lepton energy spectrum in
Ref.~\cite{gremmetal}, although the order $\Lambda_{\rm QCD}^3/m_b^3$
corrections not included in these analyses introduce large additional
uncertainties~\cite{GK,FLSmass2}.$\,\dagger$

\footnotetext{CLEO studied moments of the lepton spectrum~\cite{CLEOparams},
but the band corresponding to $\langle E_e \rangle = (1.36\pm 0.01\pm
0.02)\,$GeV on the CLEO plot cannot be reproduced using the formulae in
Ref.~\cite{Volo}.  So I consider only the result in Eq.~(\ref{cleonums}) from
Ref.~\cite{CLEOparams}.  (I thank Iain Stewart for checking this calculation.)}

In the upsilon expansion $\bar\Lambda$ is not a free parameter, so we can
determine $\lambda_1$ directly with smaller uncertainty.  Considering the
observable~\cite{gremmetal} $R_1 = \int_{1.5{\rm GeV}} E_e ({\rm d}\Gamma /
{\rm d}E_e) {\rm d}E_e \big/ \int_{1.5{\rm GeV}} ({\rm d}\Gamma / {\rm d}E_e)
{\rm d}E_e$, a fit to the same data yields~\cite{upsexp}
\begin{equation}\label{lambda1}
  \lambda_1 = (-0.27 \pm 0.10 \pm 0.04) \,{\rm GeV}^2 .
\end{equation}
This is in perfect agreement with the value of $\lambda_1$ implied by the CLEO
result for $\langle s_H - \overline{m}_D^2 \rangle$ in the upsilon expansion. 
The central value in Eq.~(\ref{lambda1}) includes corrections of order
$\alpha_s^2 \beta_0$~\cite{gremmetal2}, but the result at tree level or at
order $\alpha_s$ changes by less than $0.03\,{\rm GeV}^2$.  The first error is
dominated by $1/m_b^3$ corrections~\cite{GK} not included in
Eq.~(\ref{cleonums}).  We varied the dimension-six matrix elements between
$\pm(0.5\,{\rm GeV})^3$, and combined their coefficients in quadrature.  The
second error is from assuming a $100\,$MeV uncertainty in Eq.~(\ref{upsmass}).  

Another way to test the upsilon expansion, or determine the nonperturbative
contribution to $m_\Upsilon$, is from $\bar B\to X_s\gamma$.  Possible
contributions to the total rate from physics beyond the standard model are
unlikely to significantly affect the shape of the spectrum.  The upsilon
expansion yields parameter free predictions for moments of this distribution. 
Experimentally one needs to make a lower cut on $E_\gamma$, so it is most
convenient to study
\begin{equation}\label{moment1}
\overline{(1 - x_B)} \Big|_{x_B > 1 - \delta} = 
  { \int_{1-\delta}^1 dx_B\, (1-x_B)\, \frac{d\Gamma}{dx_B} \over
    \int_{1-\delta}^1 dx_B\, \frac{d\Gamma}{dx_B} } \,,
\end{equation}
where $x_B = 2E_\gamma/m_B$.
The parameter $\delta=1-2E_\gamma^{\rm min}/m_B$ has to satisfy $\delta >
\Lambda_{\rm QCD}/m_B$, otherwise nonperturbative effects are not under
control.  Order $\alpha_s^2\beta_0$ corrections to the photon spectrum away
from its endpoint were computed recently~\cite{LLMW}.  Fig.~1 shows the
prediction for $\overline{(1 - x_B)} |_{x_B > 1-\delta}$ as a function of
$\delta$, both at order $\epsilon$ and $(\epsilon^2)_{\rm BLM}$, neglecting
nonperturbative contributions to $m_\Upsilon$.  A $+100\,$MeV contribution
would increase $\overline{(1 - x_B)}$ by 7\%, so measuring $\overline{(1 -
x_B)}$ with such accuracy will have important implications for the physics of
quarkonia as well as for $B$ physics.

\begin{figure}[t]
\centerline{\epsfxsize=\hsize \epsffile{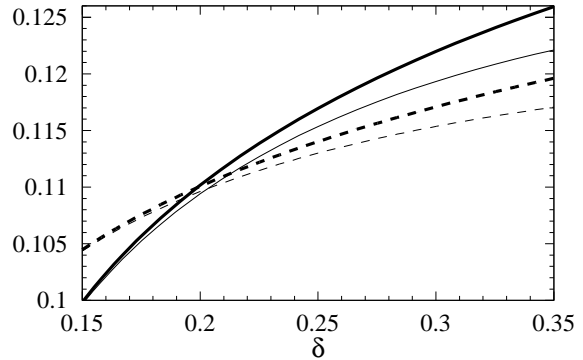}}
\caption[]{Prediction for $\overline{(1 - x_B)} |_{x_B > 1-\delta}$ in the 
upsilon expansion at order $(\epsilon^2)_{\rm BLM}$ (thick solid curve) and 
$\epsilon$ (thick dashed curve).  The thin curves show the contribution of 
the $O_7$ operator only.  (From Ref.~\cite{LLMW}.) }
\end{figure}

For $E_\gamma>2.1\,$GeV Fig.~1 gives $\overline{(1 - x_B)} = 0.111$, whereas
the central value from CLEO~\cite{CLEObsg} is around 0.093.  Interestingly,
including the CLEO data point in the $1.9\,{\rm GeV} < E_\gamma < 2.1\,$GeV
bin, the experimental central value of $\overline{(1 - x_B)}$ over the region
$E_\gamma > 1.9\,$GeV is 0.117, whereas the upsilon expansion predicts 0.120. 
Ultimately, one would like to see whether prediction and data agree over some
range of the cut $E_\gamma^{\rm min}$.  One can also evaluate $\overline{(1 -
x_B)}$ in terms of $\bar\Lambda$ and $\lambda_1$ without using the upsilon
expansion.  The CLEO data~\cite{CLEObsg} in the region $E_\gamma > 2.1\,$GeV
implies the central values $\bar\Lambda_{\alpha_s} \simeq 390\,$MeV and
$\bar\Lambda_{\alpha_s^2\beta_0} \simeq 270\,$MeV at order $\alpha_s$ and
$\alpha_s^2\beta_0$, respectively~\cite{LLMW}.  The BLM terms may not dominate
at order $\alpha_s^2$~\cite{AI}, so it is important to calculate the complete
two-loop correction to the $O_7$ contribution to $\overline{(1 - x_B)}$.

The upsilon expansion has been applied to form factor ratios in exclusive
semileptonic $B$ decays, as well as nonleptonic decays~\cite{upsexp}, where it
also improves the perturbation series (see Table 1).  However, the semileptonic
$B$ branching fraction or the average number of charm quarks in $B$ decay agree
with other predictions in the literature.  Applications of similar ideas for
$e^+e^-\to t\bar t$ are reviewed by Teubner at this Conference~\cite{TT}.

\begin{table}[t]
{\small
\[
\renewcommand{\arraystretch}{1.25}
\begin{array}{|c|r@{\hspace{0.2em}}c@{\hspace{0.2em}}
l@{\hspace{0.2em}}c@{\hspace{0.2em}}l|r@{\hspace{0.2em}}c@{\hspace{0.2em}}
l@{\hspace{0.2em}}c@{\hspace{0.2em}}l|}
\hline
{\rm Decay} & \multicolumn{10}{c|}{\rm{Expansions\ in\ terms\ of}} \\
{\rm widths} & \multicolumn{5}{c|}{m_b^{\rm pole}\ {\rm and}\ \alpha_s} & 
\multicolumn{5}{c|}{\rm{m_\Upsilon\ {\rm and}\ \epsilon  }} \\
\hline
B \to X_c e \bar\nu & 1 &-& 0.12 \epsilon &-& 0.07 \epsilon^2 &
  1 &-& 0.10 \epsilon &-& 0.03 \epsilon^2 \\
B \to X_u e \bar\nu & 1 &-& 0.17 \epsilon &-& 0.13 \epsilon^2 &
  1 &-& 0.12 \epsilon &-& 0.03 \epsilon^2 \\
B \to X_c \tau \bar\nu & 1 &-& 0.10 \epsilon &-& 0.06 \epsilon^2 & 
  1 &-& 0.07 \epsilon &-& 0.02 \epsilon^2 \\
B \to X_u \tau \bar\nu & 1 &-& 0.16 \epsilon && &
  1 &-& 0.08 \epsilon  &&\\
\hline
B \to X_{c \bar u (s+d)}& 1 &-& 0.05 \epsilon &-& 0.04 \epsilon^2 &
  1 &-& 0.03 \epsilon &-& 0.01 \epsilon^2 \\
B \to X_{c \bar c (s+d)}& 1 &+& 0.20 \epsilon &+& 0.15 \epsilon^2 &
  1 &+& 0.16 \epsilon &+& 0.07 \epsilon^2 \\
B \to X_{u \bar u (s+d)}& 1 &-& 0.10 \epsilon & & &
  1 &-& 0.05 \epsilon  &&\\
B \to X_{u \bar c (s+d)}& 1 &+& 0.09 \epsilon & &&
  1 &+& 0.11 \epsilon  &&\\
\hline
\end{array}
\]
}
\caption[]{Comparison of the perturbation series for inclusive decay rates 
using the conventional expansion and the upsilon expansion~\cite{upsexp}.  
The second order terms are the BLM parts only.}
\end{table}

\section{Conclusions}

\begin{itemize}

\item Using $m_\Upsilon$ and the upsilon expansion, i.e., assigning order
$\epsilon^n$ to the order $\alpha_s^n$ term in $B$ decay rates and
$\epsilon^{n-1}$ to the $\alpha_s^n$ term in the perturbative expression for
$m_\Upsilon$, is equivalent to using a short distance $b$ quark mass.

\item It improves the behavior of perturbation series for inclusive $B$ decays,
and eliminates $m_b$ altogether from the theoretical predictions in favor of
$m_\Upsilon$ in a simple and consistent manner.  

\item It may lead to smaller nonperturbative effects (to the extent this is
reflected in the behavior of perturbation series).

\item The biggest uncertainty is the nonperturbative contribution to
$m_\Upsilon$ unrelated to the quark mass.  It will be possible to estimate /
constrain this from data in the future.

\end{itemize}

\section*{Acknowledgements}

I thank Andre Hoang, Mike Luke, Aneesh Manohar, and Mark Wise for very
enjoyable collaborations on the topics described in this talk.  Fermilab is
operated by Universities Research Association, Inc., under DOE contract
DE-AC02-76CH03000.

\end{document}